\documentstyle[11pt]{article}
\def\fnote#1#2{\begingroup\def\thefootnote{#1}\footnote{#2}\addtocounter
{footnote}{-1}\endgroup}

\begin{document}

\hfill{UTTG-09-09}

\hfill{TCC-028-09}

\vspace{36pt}

\begin{center}
{\large {\bf {Effective Field Theory, Past and Future}}}

\vspace{36pt}
Steven Weinberg\fnote{*}{Electronic address:
weinberg@physics.utexas.edu}\\
{\em Theory Group, Department of Physics, University of
Texas\\
Austin, TX, 78712}

\vspace{30pt}

\noindent
{\bf Abstract}
\end{center}
\noindent
This is a written version of the opening talk at the 6th International Workshop on Chiral Dynamics, at the University of Bern, Switzerland, July 6, 2009, to be published in the proceedings of  the Workshop.  In it, I reminisce about the early development of effective field theories of the strong interactions, comment briefly on some other applications of effective field theories, and then take up the idea that the Standard Model and General Relativity are the leading terms in an effective field theory.  Finally, I cite recent calculations that suggest that the effective field  theory of gravitation and matter is asymptotically safe.

\vfill

\pagebreak

I have been asked by the organizers of this meeting to  ``celebrate 30 years'' of a paper\footnote{S. Weinberg, Physica A96, 327 (1979).} on effective field theories that I wrote in 1979.  I am quoting this request at the outset, because in the first half of this talk I will be reminiscing about my own work on effective field theories, leading up to this 1979 paper.  I think it is important to understand how confusing these things seemed back then, and no one knows better than I do how confused I was.  But I am sure that many in this audience know more than I do about the applications of effective field theory to the strong interactions since 1979, so I will mention only some early applications to strong interactions and a few applications to other areas of physics.   I will then describe how we have come to think that our most fundamental theories, the Standard Model and General Relativity, are  the leading terms in an effective field theory.  Finally, I will report on recent work of others that lends support to a suggestion that this effective field theory may actually be a fundamental theory, valid at all energies.
  
It all started with current algebra.  As everyone  knows, in 1960 Yoichiro Nambu had the idea that the axial vector current of beta decay could be considered to be conserved in the same limit that the pion, the lightest hadron, could be considered massless.\footnote{Y. Nambu, Phys. Rev. Lett. 4, 380 (1960).}  This assumption would follow if the axial vector current was associated with a spontaneously broken approximate symmetry, with the pion playing the role of a Goldstone boson.\footnote{J. Goldstone, Nuovo Cimento 9, 154 (1961); Y. Nambu and G. Jona-Lasinio, Phys. Rev. 122, 345 (1961); J. Goldstone, A. Salam, and S. Weinberg, Phys. Rev. 127, 965 (1962).  }  Nambu used this idea to explain the success of the Goldberger-Treiman formula\footnote{M. L. Goldberger and S. B. Treiman, Phys. Rev. 111, 354 (1956).} for the pion decay amplitude, and with his collaborators he was able to derive formulas for the rate of emission of a single low energy pion in various collisions.\footnote{Y. Nambu and D. Lurie, Phys Rev. 125, 1429 (1962); Y. Nambu and E. Shrauner, Phys. Rev. 128, 862 (1962).}  In this work it was not necessary to assume anything about the nature of the broken symmetry -- only that there was some approximate symmetry responsible for the approximate conservation of the axial vector current and the approximate masslessness of the pion.  But to deal with processes involving more than one pion, it was necessary  to use not only the approximate conservation of the current but also the current commutation relations,   which of course do depend on the underlying broken symmetry.  The technology of using these properties of the currents, in which one does not use any specific Lagrangian for the strong interactions, became known as current algebra.\footnote{The name may be due to Murray Gell-Mann.  The current commutation relations were given in M. Gell-Mann, Physics 1, 63 (1964).}  It scored a dramatic success in the derivation of the Adler-Weisberger sum rule\footnote{S. L. Adler, Phys. Rev. Lett. 14, 1051 (1965); Phys. Rev. 140, B736  (1965); W. I. Weisberger, Phys. Rev. Lett. 14, 1047 (1965).} for the axial vector beta decay coupling constant $g_A$, which showed that the current commutation relations are those of $SU(2)\times SU(2)$.  

When I started  in the mid-1960s to work on current algebra, I had the feeling that, despite the success of the Goldberger-Treiman relation and the Adler-Weisberger sum rule, there was then rather too much emphasis on the role that the axial vector current plays in weak interactions.  After all, even if there were no weak interactions, the fact that the strong interactions have an approximate but spontaneously broken $SU(2)\times SU(2)$ symmetry would be a pretty important piece of information about the strong interactions.\footnote{I emphasized this point in my rapporteur's talk on current algebra at the 1968 ``Rochester'' conference; see {\em Proceedings of the 14th International Conference on High-Energy Physics}, p. 253.}  To demonstrate the point, I was able to use current algebra to derive successful formulas for the pion-pion and pion-nucleon scattering lengths.\footnote{S. Weinberg, Phys. Rev. Lett. 17, 616 (1966).  The pion-nucleon scattering lengths were calculated independently by Y. Tomozawa, Nuovo Cimento 46A, 707 (1966).} When combined with a well-known dispersion relation\footnote{M. L. Goldberger, Y. Miyazawa, and R. Oehme, Phys. Rev. 99, 986 (1955).} and the Goldberger-Treiman relation, these formulas for the pion--nucleon scattering lengths turned out to imply the Adler-Weisberger sum rule.  

In 1966 I turned to the problem of calculating the rate of processes in which arbitrary numbers of low energy massless pions are emitted in the collision of other hadrons.  This was not a problem that urgently needed to be solved.  I was interested in it because a  year earlier I had worked out simple formulas for the rate of emission of arbitrary numbers of soft gravitons or photons in any collision,\footnote{S. Weinberg, Phys. Rev. 140, B516 (1965).} and I was curious whether  anything equally simple could be said about  soft pions.  Calculating the amplitude for emission of several soft pions by use of the technique of current algebra turned out to be fearsomely complicated; the non-vanishing commutators of the currents associated with the soft pions  prevented the derivation of anything as simple as the results for soft photons or gravitons, except in the special case in which all pions have the same charge.\footnote{ S. Weinberg, Phys. Rev. Lett. 16, 879 (1966).}    

Then some time late in 1966 I was sitting at the counter of a caf\'{e} in Harvard Square,  scribbling on a napkin the amplitudes I had found for emitting two or three soft pions in nucleon collisions, and it suddenly occurred to me that these results looked very much like what would be given by lowest order Feynman diagrams in a quantum field theory in which pion lines are emitted from the external nucleon lines, with a Lagrangian in which the nucleon interacts with one, two, and more pion fields.  Why should this be?  Remember, this was a time when theorists had pretty well given up the idea of applying specific quantum field theories  to the strong interactions, because there was no reason to trust the lowest order of perturbation theory, and no way to sum the perturbation series.  What was popular was to exploit tools such as current algebra and dispersion relations that did not rely on assumptions about particular Lagrangians.  

The best explanation  that I could give then for the field-theoretic appearance of the current algebra results was that these results for the emission of $n$ soft pions in nucleon collisions are of the minimum order, $G_\pi^n$,  in the pion-nucleon coupling constant $G_\pi$,  except that one had to use the exact values for the collision amplitudes without soft pion emission, and  divide by factors of the axial vector coupling constant $g_A\simeq 1.2$ in appropriate places.  Therefore any Lagrangian that satisfied the axioms of current algebra would have to give the same answer as current algebra in lowest order perturbation theory, except that it would have to be a field theory in which soft pions were emitted only from external lines of the diagram for the nucleon collisions, for only then would one know how to put in the correct factors of $g_A$ and the correct nucleon collision amplitude.  

The time-honored renormalizable theory of nucleons and pions with conserved currents that satisfied the assumptions of current algebra was the ``linear $\sigma$-model,''\footnote{J. Bernstein, S. Fubini, M. Gell-Mann, and W. Thirring, Nuovo Cimento 17, 757 (1960); M. Gell-Mann and M. L\'{e}vy, Nuovo Cimento 16, 705 (1960); K. C. Chou, Soviet Physics JETP 12, 492 (1961).  This theory, with the inclusion of a symmetry-breaking term proportional to the $\sigma$ field, was intended to provide an illustration of a ``partially conserved axial current,'' that is, one whose divergence is proportional to the pion field.  } with Lagrangian (in the limit of exact current conservation):
\begin{eqnarray}
{\cal L}&=&-\frac{1}{2}\left[\partial_\mu\vec{\pi}\cdot\,\partial^\mu\vec{\pi}+\partial_\mu\sigma\,\partial^\mu\sigma\right]\nonumber\\
&&-\frac{m^2}{2}\Big(\sigma^2+{\vec{\pi}}^2\Big)-\frac{\lambda}{4}\Big(\sigma^2+{\vec{\pi}}^2\Big)^2\nonumber\\
&&-\bar{N}\gamma^\mu\partial_\mu N-G_\pi\bar{N}\Big(\sigma+2i\gamma_5\vec{\pi}
\cdot\vec{t}\Big)N\;,
\end{eqnarray}
where $N$, $\vec{\pi}$, and $\sigma$ are the fields of the nucleon doublet, pion triplet, and a scalar singlet, and $\vec{t}$ is the nucleon isospin matrix (with $\vec{t}^2=3/4$).  This Lagrangian has an  $SU(2)\times SU(2)$ symmetry (equivalent as far as current commutation relations are concerned to an $SO(4)$ symmetry), that is spontaneously broken for $m^2<0$ by the expectation value of the $\sigma$ field, given in lowest order by $<\sigma>=F/2\equiv \sqrt{-m^2/\lambda}$, which also gives the nucleon a lowest order mass $2G_\pi F$.  But with a Lagrangian of this form soft pions could be emitted from internal as well as external lines of the graphs for the nucleon collision itself, and there would be no way to evaluate the pion emission amplitude without having to sum over the infinite number of graphs for the nucleon collision amplitude.  

To get around this obstacle, I used the chiral $SO(4)$ symmetry to rotate the chiral four-vector into the fourth direction
\begin{equation}
\Big(\vec{\pi},\sigma\Big)\mapsto \Big(0,\sigma'\Big)\;,~~~~~~~\sigma'=\sqrt{\sigma^2+{\vec{\pi}}^2}\;,
\end{equation}
with a corresponding chiral transformation $N\mapsto N'$ of the nucleon doublet.
The chiral symmetry of the Lagrangian would result in the pion disappearing from the Lagrangian, except that the matrix of the rotation (2)  necessarily, like the fields, depends on spacetime position, while the theory is only invariant under spacetime-{\em independent} chiral rotations.  The pion field thus reappears as a parameter in the $SO(4)$ rotation (2), which could conveniently be taken as
\begin{equation}
\vec{\pi}'\equiv F\vec{\pi}/[\sigma+\sigma']\;.
\end{equation}
But the rotation parameter  $\vec{\pi}'$ would not appear in the transformed Lagrangian if it were independent of the spacetime coordinates, so wherever it appears it must be accompanied with at least one derivative.  This derivative produces a factor of pion four-momentum in the pion emission amplitude, which would suppress the amplitude for emitting soft pions, if this factor were not compensated by the pole in the nucleon propagator of an external nucleon line to which the pion emission vertex is attached. Thus, with the Lagrangian in this form,  pions of small momenta can only be emitted from external lines of a nucleon collision amplitude.  This is what I needed.  

Since $\sigma'$ is chiral-invariant, it plays no role in maintaining the chiral invariance of the theory, and  could therefore be replaced with its lowest-order expectation value $F/2$.  The transformed Lagrangian (now dropping primes) is then
\begin{eqnarray}
{\cal L}&=&-\frac{1}{2}\left[1+\frac{{\vec{\pi}}^2}{F^2}\right]^{-2}\partial_\mu\vec{\pi}\cdot \partial^\mu\vec{\pi}
\nonumber\\&&-\bar{N}\Bigg[\gamma^\mu\partial_\mu+G_\pi F/2\nonumber\\&&~~~+i\gamma^\mu \left[1+\frac{{\vec{\pi}}^2}{F^2}\right]^{-1}\left[\frac{2}{F}\gamma_5
\vec{t}\cdot\partial_\mu\vec{\pi}+\frac{2}{F^2}\vec{t}\cdot (\vec{\pi}\times \partial_\mu\vec{\pi})\right]\Bigg]N\;.
\end{eqnarray}
In order to reproduce the results of current algebra, it is only necessary to  identify $F$ as the pion decay amplitude $F_\pi\simeq 184$ MeV, replace the term $G_\pi F/2$ in the nucleon bilinear with the actual nucleon mass $m_N$ (given by the Goldberger--Treiman relation as $G_\pi F_\pi/2g_A$), and replace the pseudovector pion-nucleon coupling $1/F$ with its actual value $G_\pi/2m_N=g_A/F_\pi$.  This gives an effective Lagrangian  
\begin{eqnarray}
{\cal L}_{\rm eff}&=&-\frac{1}{2}\left[1+\frac{{\vec{\pi}}^2}{F_\pi^2}\right]^{-2}\partial_\mu\vec{\pi}\cdot \partial^\mu\vec{\pi}
\nonumber\\&&-\bar{N}\Bigg[\gamma^\mu\partial_\mu+m_N\nonumber\\&&~~~+i\gamma^\mu \left[1+\frac{{\vec{\pi}}^2}{F_\pi^2}\right]^{-1}\left[\frac{G_\pi}{m_N}\gamma_5
\vec{t}\cdot\partial_\mu\vec{\pi}+\frac{2}{F_\pi^2}\vec{t}\cdot (\vec{\pi}\times \partial_\mu\vec{\pi})\right]\Bigg]N\;.
\end{eqnarray}
To take account of the finite pion mass, the linear sigma model also includes a chiral-symmetry breaking perturbation proportional to $\sigma$.  Making the chiral rotation (2), replacing $\sigma'$ with the constant $F/2$, and adjusting the coefficient of this term to give the physical pion mass $m_\pi$ gives a chiral symmetry breaking term
\begin{equation}
\Delta{\cal L}_{\rm eff}=-\frac{1}{2}\left[1+\frac{{\vec{\pi}}^2}{F_\pi^2}\right]^{-1}m_\pi^2\,{\vec{\pi}}^2\;.
\end{equation}
Using ${\cal L}_{\rm eff}+\Delta {\cal L}_{\rm eff}$ in lowest order perturbation theory, I found  the same results for low-energy pion-pion and pion-nucleon scattering that I had  obtained earlier with much greater difficulty by the methods of current algebra.

A few months after this work,  Julian Schwinger remarked to me that it should be possible to skip this complicated derivation, forget all about the linear $\sigma$-model, and instead infer the structure of the Lagrangian directly from the non-linear chiral transformation properties of the pion field appearing in (5).\footnote{For Schwinger's own development of this idea, see J. Schwinger, Phys. Lett. 24B, 473 (1967).  It is interesting that in deriving the effective field theory of goldstinos in supergravity theories, it is much more transparent to start with a theory with linearly realized supersymmetry and impose constraints analogous to setting $\sigma'=F/2$, than to work from the beginning with supersymmetry realized non-linearly, in analogy to Eq.~(7); see Z. Komargodski and N. Seiberg, to be published.}  It was a good idea.  I spent the summer of 1967 working out these transformation properties, and what they imply for the structure of the Lagrangian.\footnote{S. Weinberg, Phys. Rev. 166, 1568 (1968).}  It turns out that if we require that the pion field  has the usual linear transformation under $SO(3)$ isospin rotations (because isospin symmetry is supposed to be not spontaneously broken), then there is a {\em unique} $SO(4)$ chiral transformation that takes the pion field into a function of itself --- unique, that is, up to possible redefinition of the field.  For an infinitesimal $SO(4)$ rotation by an angle $\epsilon$ in the $a4$ plane (where $a=1,2,3$), the pion field $\pi_b$ (labelled with a prime in Eq.~(3)) changes by an amount 
\begin{equation}
\delta_a\pi_b= -i\epsilon F_\pi\left[\frac{1}{2}\left(1-\frac{{\vec{\pi}}^2}{F_\pi^2}\right)\delta_{ab}+\frac{\pi_a\pi_b}{F_\pi^2}\right]\;.
\end{equation}
Any other field $\psi$, on which isospin rotations act with a matrix $\vec{t}$ , is changed by an infinitesimal chiral rotation in the $a4$ plane by an amount  
\begin{equation}
\delta_a\psi=\frac{\epsilon}{F_\pi}\Big(\vec{t}\times \vec{\pi}\Big)_a\psi\;.
\end{equation}
This is just an ordinary, though position-dependent, isospin rotation, so a non-derivative isospin-invariant term  in the Lagrangian that does not involve pions, like the nucleon mass term $-m_N \bar{N}N$, is automatically chiral-invariant.  The terms in Eq.~(5):
\begin{equation}
-\bar{N}\Bigg[\gamma^\mu\partial_\mu+\frac{2i}{F_\pi^2}\gamma^\mu \left[1+\frac{{\vec{\pi}}^2}{F_\pi^2}\right]^{-1}\vec{t}\cdot (\vec{\pi}\times \partial_\mu\vec{\pi})\Bigg]N \;,
\end{equation}
and 
\begin{equation}
-i\frac{G_\pi}{m_N} \left[1+\frac{{\vec{\pi}}^2}{F_\pi^2}\right]^{-1}\bar{N}\gamma^\mu\gamma_5
\vec{t}\cdot\partial_\mu\vec{\pi}N\;,
\end{equation}
are simply proportional to the most general chiral-invariant nucleon--pion interactions with a single spacetime derivative.  The coefficient of the term (9) is fixed by the condition that $N$ should be canonically normalized, while the coefficient of (10) is chosen to agree with the conventional definition of the 
pion-nucleon coupling $G_\pi$, and is not directly constrained by chiral symmetry.   The term 
\begin{equation}
-\frac{1}{2}\left[1+\frac{{\vec{\pi}}^2}{F^2}\right]^{-2}\partial_\mu\vec{\pi}\cdot \partial^\mu\vec{\pi}
\end{equation}
is proportional to the most general chiral invariant quantity involving the pion field and no more than two spacetime derivatives, with a coefficient fixed by the condition that $\vec{\pi}$ should be canonically normalized.  
The chiral symmetry breaking term (6) is the most general function of the pion field without derivatives that transforms as the fourth component of a chiral four-vector.  None of this relies on the methods of current algebra, though one can use the Lagrangian (5) to calculate the Noether current corresponding to chiral transformations, and recover the Goldberger-Treiman relation in the form $g_A=G_\pi F_\pi/2m_N$.  

This sort of direct analysis was subsequently extended by Callan, Coleman, Wess, and Zumino to the transformation and interactions of the Goldstone boson fields associated with the spontaneous breakdown of any Lie group $G$ to any subgroup $H$.\footnote{S. Coleman, J. Wess, and B. Zumino, Phys. Rev. 177, 2239(1969); C. G. Callan, S. Coleman, J. Wess, and B. Zumino, Phys. Rev. 177, 2247(1969).}   Here, too, the transformation of the Goldstone boson fields is unique, up to a redefinition of the fields, and the transformation of other fields under $G$ is uniquely determined by their transformation under the unbroken subgroup $H$.  It is straightforward to work out the rules for using these ingredients to construct effective Lagrangians that are invariant under $G$ as well as $H$.\footnote{There is a complication.  In some cases,  such as $SU(3)\times SU(3)$ spontaneously broken to $SU(3)$, fermion loops produce $G$-invariant terms in the action that are not the integrals of $G$-invariant terms in the Lagrangian density; see J. Wess and B. Zumino, Phys. Lett. 37B, 95 (1971); E. Witten, Nucl. Phys. B223, 422 (1983).  The most general such terms in the action, whether or not produced by fermion loops, have been cataloged by E. D'Hoker and S. Weinberg, Phys. Rev. D50, R6050 (1994).  It turns out that for $SU(N)\times SU(N)$ spontaneously broken to the diagonal $SU(N)$, there is just one such term for $N\geq 3 $, and none for $N=1$ or $2$.  For $N=3$, this term is the one found by Wess and Zumino.}    Once again, the key point is that the invariance of the Lagrangian under $G$ would eliminate all presence of the Goldstone boson field in the Lagrangian if the field were spacetime-independent, so wherever functions of this field appear in the Lagrangian they are always accompanied with at least one spacetime derivative.  

In the following years, effective Lagrangians with spontaneously broken $SU(2)\times SU(2)$ or $SU(3)\times SU(3)$ symmetry were widely used in lowest-order perturbation theory to make predictions about low energy pion and kaon interactions.\footnote{For reviews, see S. Weinberg, in {\em Lectures on Elementary Particles and Quantum Field Theory --- 1970 Brandeis University Summer Institute in Theoretical Physics, Vol. 1}, ed. S. Deser, M. Grisaru, and H. Pendleton (The M.I.T. Press, Cambridge, MA, 1970); B. W. Lee,  {\em Chiral Dynamics} (Gordon and Breach, New York, 1972).}  But during this period, from the late 1960s to the late 1970s,  like many other particle physicists I was chiefly concerned with developing and testing the Standard Model of elementary particles.  As it happened, the Standard Model did much to clarify the basis for chiral symmetry.  Quantum chromodynamics with $N$ light quarks is automatically invariant under a $SU(N)\times SU(N)$ chiral symmetry,\footnote{For a while it was not clear why there was not also a chiral $U(1)$ symmetry, that would also be broken in the Lagrangian only by the quark masses, and would either lead to a parity doubling of observed hadrons, or to a new light pseudoscalar neutral meson, both of which possibilities were experimentally ruled out.  It was not until 1976 that `t Hooft pointed out that the effect of triangle anomalies in the presence of instantons produced an intrinsic violation of this unwanted chiral $U(1)$ symmetry; see G. `t Hooft, Phys. Rev. D14, 3432 (1976).}  broken in the Lagrangian only by quark masses, and the electroweak theory tells us that the currents of this symmetry (along with the  quark number currents) are just those to which the $W^\pm$, $Z^0$, and photon are coupled.

During this whole period, effective field theories appeared as only a device for more easily reproducing the results of current algebra.  It was difficult to take them seriously as dynamical theories, because the derivative couplings that made them useful in the lowest order of perturbation theory also made them nonrenormalizable, thus apparently closing off the possibility of using these theories in higher order.  

My thinking about this began to change in 1976.  I was invited to give a series of lectures at Erice that summer, and took the opportunity to learn the theory of critical phenomena by giving lectures about it.\footnote{S. Weinberg, ``Critical Phenomena for Field Theorists,'' in {\em Understanding the Fundamental Constituents of Matter}, ed. A. Zichichi (Plenum Press, New York, 1977).}  In preparing these lectures, I was struck by Kenneth Wilson's device of ``integrating out'' short-distance degrees of freedom by introducing a variable ultraviolet cutoff, with the bare couplings given a cutoff dependence that guaranteed that physical quantities are cutoff independent.  Even if the underlying theory is renormalizable, once a finite cutoff is introduced it becomes necessary to introduce every possible interaction, renormalizable or not, to keep physics strictly cutoff independent.  From this point of view, it doesn't make much difference whether the underlying theory is renormalizable or not.  Indeed, I realized that even without a cutoff, as long as every term allowed by symmetries is included in the Lagrangian, there will always be a counterterm available to absorb every possible ultraviolet divergence by renormalization of the corresponding coupling constant.  Non-renormalizable theories, I realized, are just as renormalizable as renormalizable theories.  

This opened the door to the consideration of a Lagrangian containing terms like (5)  as the basis for a legitimate dynamical theory, not limited to the tree approximation, provided one adds  every one of the infinite number of  other, higher-derivative,  terms  allowed by chiral symmetry.\footnote{I thought it appropriate to publish this in a festschrift for Julian Schwinger; see footnote 1.}   But for this to be useful, it is necessary that in some sort of perturbative expansion, only a finite number of terms in the Lagrangian can appear in each order of perturbation theory.

In chiral dynamics, this perturbation theory is provided by an expansion in powers of small momenta and pion masses.   At momenta of order $m_\pi$, the number $\nu$ of factors of momenta or $m_\pi$ contributed by a diagram with $L$ loops, $E_N$ external nucleon lines, and $V_i$ vertices of type $i$, for any reaction among pions and/or nucleons, is
\begin{equation}
\nu=\sum_i V_i\left(d_i+\frac{n_i}{2}+m_i-2\right)+2L+2-\frac{E_N}{2}\;,
\end{equation}
where $d_i$, $n_i$, and $m_i$ are respectively the numbers of derivatives, factors of nucleon fields, and factors of pion mass (or more precisely, half the number of factors of $u$ and $d$ quark masses) associated with vertices of type $i$.  As a consequence of chiral symmetry, the minimum possible value of $d_i+n_i/2+m_i$ is 2, so the leading diagrams for small momenta are those with $L=0$ and any number of interactions with $d_i+n_i/2+m_i=2$, which are  the ones given in Eqs.~(5) and (6).  To next order in momenta, we may include tree graphs with any number of vertices with $d_i+n_i/2+m_i=2$ and just one vertex with $d_i+n_i/2+m_i=3$ (such as the so-called $\sigma$-term).  To next order, we include any number of vertices with $d_i+n_i/2+m_i=2$, plus either a single loop, or a single vertex with $d_i+n_i/2+m_i=4$ which provides a counterterm for the infinity in the loop graph, or two vertices with $d_i+n_i/2+m_i=3$.  And so on.  Thus one can generate a power series in momenta and $m_\pi$, in which only a few new constants need to be introduced at each new order.  As an explicit example of this procedure, I calculated the one-loop corrections to pion--pion scattering in the limit of zero pion mass, and of course I found the sort of corrections required to this order by unitarity.\footnote{Unitarity corrections to soft-pion results of current algebra had been considered earlier by H. Schnitzer, Phys. Rev. Lett. 24, 1384 (1970); Phys. Rev. D2, 1621 (1970); L.-F. Li and H. Pagels, Phys. Rev. Lett. 26, 1204 (1971); Phys. Rev. D5, 1509 (1972); P. Langacker and H. Pagels, Phys. Rev. D8, 4595 (1973).}

But even if this procedure gives well-defined finite results, how do we know they are true?  It  would be extraordinarily difficult to justify any calculation involving loop graphs using current algebra.  For me in 1979, the answer involved a radical reconsideration of the nature of quantum field theory.  From its beginning in the late 1920s, quantum field theory had been regarded as the application of quantum mechanics to fields that are among the fundamental constituents of the universe --- first the electromagnetic field, and later the electron field and fields for other known ``elementary'' particles.  In fact, this became a working definition of an elementary particle --- it is a particle with its own field.                                                                                  But  for years in  teaching courses on quantum field theory I had emphasized that the description of nature by quantum field theories is inevitable, at least in theories with a finite number of particle types, once one assumes the principles of relativity and quantum mechanics, plus the cluster decomposition principle, which requires that distant experiments have uncorrelated results.  So I began to think that although specific quantum field theories may have a content that goes beyond these general principles, quantum field theory itself does not.  I offered this in my 1979 paper as  what Arthur Wightman would call a folk theorem: ``if one writes down the most general possible Lagrangian, including {\em all} terms consistent with assumed symmetry principles, and then calculates matrix elements with this Lagrangian to any given order of perturbation theory, the result will simply be the most general possible $S$-matrix consistent with  perturbative unitarity,  analyticity, cluster decomposition, and the assumed symmetry properties.''  So current algebra wasn't needed.  

There was an interesting irony in this.  I had been at Berkeley from 1959 to 1966, when Geoffrey Chew and his collaborators were elaborating a program for calculating $S$-matrix elements for strong interaction processes by the use of unitarity, analyticity, and Lorentz invariance, without reference to quantum field theory.  I found it an attractive philosophy, because it relied only on a minimum of principles, all well established.  Unfortunately, the $S$-matrix theorists were never able to develop a reliable method of calculation, so I worked instead on other things, including current algebra.  Now in 1979 I realized that the assumptions of $S$-matrix theory, supplemented by chiral invariance, were indeed all that are needed at low energy, but the most convenient way of implementing  these assumptions in actual calculations was by good old quantum field theory, which the $S$-matrix theorists had hoped to supplant.

After 1979,  effective field theories were applied to strong interactions in 
 work by Gasser, Leutwyler, Meissner, and many other theorists.  My own contributions to this work were limited to two areas --- isospin violation, and nuclear forces.

At first in the development of chiral dynamics there had been a tacit assumption  that isotopic spin symmetry was a better approximate symmetry than chiral $SU(2)\times SU(2)$, and that the Gell-Mann--Ne'eman $SU(3)$ symmetry was a better approximate symmetry than 
chiral $SU(3)\times SU(3)$.  This assumption became untenable with the calculation of quark mass ratios from the measured pseudoscalar meson masses.\footnote{S. Weinberg, contribution to a festschrift for I. I. Rabi, Trans. N. Y. Acad. Sci. 38, 185 (1977).}  It turns out that the $d$ quark mass is almost twice the $u$ quark mass, and the $s$ quark mass is very much larger than either.  As a consequence of the inequality of $d$ and $u$ quark masses, chiral $SU(2)\times SU(2)$ is broken in the Lagrangian of quantum chromodynamics not only by the fourth component  of a chiral four-vector, as in (6), but also by the third component of a different chiral four-vector proportional to $m_u-m_d$ (whose fourth component is a pseudoscalar).  There is no function of the  pion field alone, without derivatives, with the latter transformation property, which is why pion--pion scattering and the pion masses are described by (6) and the first term in (5) in leading order, with no isospin breaking aside of course from that due to electromagnetism.  But there are non-derivative corrections to pion--nucleon interactions,\footnote{S. Weinberg, in {\em Chiral Dynamics: Theory and Experiment --- Proceedings of the Workshop Held at MIT, July 1994} (Springer-Verlag, Berlin, 1995).  The terms in Eq.~(13) that are odd in the pion field are given in Section 19.5 of S. Weinberg, {\em The Quantum Theory of Fields}, Vol. II (Cambridge University Press, 1996)} which at momenta of order $m_\pi$  are suppressed relative to the derivative coupling terms in (5) by just one factor of $m_\pi$ or momenta:  
\begin{eqnarray}
\Delta'{\cal L}_{\rm eff}&=&-\frac{A}{2}\left(\frac{1-\pi^2/F_\pi^2}{1+\pi^2/F_\pi^2}\right)\,\bar{N}N \nonumber\\
&& -B\left[\bar{N}t_3N-\frac{2}{F_\pi^2}\left(\frac{\pi_3}{1+\pi^2/F_\pi^2}\right)\,\bar{N}\vec{t}\cdot\vec{\pi}N\right]\nonumber\\&& -\frac{iC}{1+\vec{\pi}^2/F_\pi^2}\bar{N}\gamma_5 \vec{\pi}\cdot\vec{t}N\nonumber\\&&
-\frac{iD\pi_3}{1+\vec{\pi}^2/F_\pi^2}\bar{N}\gamma_5 N\;,
\end{eqnarray}
where $A$ and $C$ are proportional to $m_u+m_d$, and $B$ and $D$ are proportional to $m_u-m_d$, with $B\simeq -2.5$ MeV.
The $A$ and $B$ terms contribute isospin conserving and violating terms to the so-called $\sigma$-term in pion nucleon scattering.  

My work on nuclear forces began one day in 1990 while I was lecturing to a graduate class at Texas.  I derived Eq.~(12) for the class, and showed how the interactions in the leading tree graphs with $d_i+n_i/2+m_i=2$ were just those given  here in Eqs.~(5) and (6).  Then, while I was standing at the blackboard, it suddenly occurred to me that there was one other term with $d_i+n_i/2+m_i=2$ that I had never previously considered: an interaction with no factors of pion mass and no derivatives (and hence, according to chiral symmetry, no pions), but {\em four} nucleon fields --- that is, a sum of Fermi interactions $(\bar{N}\Gamma N)(\bar{N}\Gamma' N)$, with any matrices $\Gamma$ and $\Gamma'$ allowed by Lorentz invariance, parity conservation, and isospin conservation.  This is just the sort of ``hard core'' nucleon--nucleon interaction that nuclear theorists had long known has to be added to the pion-exchange term in  theories of nuclear force.  But there is a complication --- in graphs for nucleon--nucleon scattering at low energy, two-nucleon intermediate states make a large contribution that invalidates the sort of power-counting that justifies the use of the effective Lagrangian (5), (6) in processes involving only pions, or one low-energy nucleon plus pions.  So it is necessary to apply the effective Lagrangian, including the terms $(\bar{N}\Gamma N)(\bar{N}\Gamma' N)$ along with the terms (5) and (6), to the two-nucleon irreducible nucleon--nucleon potential, rather than directly to the scattering amplitude.\footnote{S. Weinberg, Phys. Lett. B251, 288 (1990); Nucl. Phys. B363, 3 (1991); Phys. Lett. B295, 114 (1992).}  This program was initially carried further by Ordo\~{n}ez, van Kolck, Friar, and their collaborators,\footnote{C. Ordo\~{n}ez and U. van Kolck, Phys. Lett. B291, 459 (1992); C. Ordo\~{n}ez. L. Ray,  and U. van Kolck, Phys. Rev. Lett. 72, 1982 (1994);  U. van Kolck, Phys. Rev. C49, 2932 (1994);  U. van Kolck, J. Friar, and T. Goldman, Phys. Lett. B 371, 169 (1996); C. Ordo\~{n}ez, L. Ray,  and U. van Kolck, Phys. Rev. C 53, 2086 (1996);  C. J. Friar, Few-Body Systems Suppl. 99, 1 (1996).}
and eventually by several  others.  

The advent of effective field theories generated changes in point of view and suggested new techniques of calculation that propagated out to numerous areas of physics, some quite far removed from particle physics.
Notable here is the  use of the power-counting arguments of effective field theory to justify the approximations made in the BCS theory of superconductivity.\footnote{G. Benfatto and G. Gallavotti,  J. Stat. Phys.
 59, 541 (1990);  Phys. Rev. 42, 9967
(1990); J. Feldman and E. Trubowitz,  Helv. Phys. Acta
 63, 157 (1990);  64, 213 (1991);  65, 679
(1992); R. Shankar,  Physica A177, 530 (1991);
 Rev. Mod. Phys. 66, 129 (1993); J. Polchinski,
in {\em Recent Developments in Particle Theory, Proceedings
of the 1992 TASI}, eds. J. Harvey and J. Polchinski (World
Scientific, Singapore, 1993); S. Weinberg, Nucl. Phys.  B413, 567 (1994).}  Instead of counting powers of small momenta, one must count powers of the departures of  momenta from the Fermi surface.  Also, general features of theories of inflation have been clarified by re-casting these theories as effective field theories of the inflaton and gravitational fields.\footnote{C. Cheung, P. Creminilli, A. L. Fitzpatrick, J. Kaplan, and L. Senatore, J. High Energy Physics 0803, 014 (2008); S. Weinberg, Phys. Rev. D {\bf 73}, 123541 (2008).}

Perhaps the most important lesson from chiral dynamics was that we should keep an open mind about renormalizability.  The  renormalizable Standard Model of elementary particles may itself be just the first term in an effective field theory that contains every possible interaction allowed by Lorentz invariance and the $SU(3)\times SU(2)\times U(1)$ gauge symmetry, only with the non-renormalizable terms suppressed by negative powers of some very large mass $M$, just as the terms in chiral dynamics with more derivatives than in Eq.~(5) are suppressed by negative powers of $2\pi F_\pi\approx m_N$.  One indication  that there is a large mass scale in some theory underlying the Standard Model is the well-known fact that the three (suitably normalized) running gauge couplings of $SU(3)\times SU(2) \times U(1)$ become equal at an energy of the order of $10^{15}$ GeV (or, if supersymmetry is assumed,  $2\times 10^{16}$ GeV, with better convergence of the couplings.)  

In 1979 papers by  Frank Wilczek and Tony Zee\footnote{F. Wilczek and A. Zee, Phys. Rev. Lett. 43, 1571 (1979).} and me\footnote{S. Weinberg, Phys. Rev. Lett. 43, 1566 (1979).} independently pointed out that, while the renormalizable terms of the Standard Model cannot violate baryon or lepton conservation,\footnote{This is not true if the effective theory contains fields for the squarks and sleptons of supersymmetry.  However, there are no renormalizable baryon or lepton violating terms  in ``split supersymmetry'' theories, in which the squarks and sleptons are superheavy, and only the gauginos and perhaps higgsinos survive to ordinary energies; see N. Arkani-Hamed and S. Dimopoulos, JHEP {\bf 0506}, 073 (2005); G. F. Giudice and A. Romanino, Nucl. Phys. B {\bf 699}, 65 (2004); N. Arkani-Hamed, S. Dimopoulos, G. F. Giudice, and A. Romanino, Nucl. Phys. B {\bf 709}, 3 (2005); A. Delgado and G. F. Giudice, Phys. Lett. B627, 155 (2005).} this is not true of the higher non-renormalizable terms.  In particular, four-fermion terms can generate a proton decay into antileptons, though not into leptons, with an amplitude suppressed on dimensional grounds by a factor $M^{-2}$.  The conservation of baryon and lepton number in observed physical processes thus may be an accident, an artifact of the necessary simplicity of the leading renormalizable $SU(3)\times SU(2)\times U(1)$-invariant interactions.  I also noted at the same time that interactions between a pair of lepton doublets and a pair of scalar doublets can generate a neutrino mass, which is suppressed only by a factor $M^{-1}$, and that therefore with a reasonable estimate of $M$ could produce observable neutrino oscillations.  The subsequent confirmation of neutrino oscillations lends support to the view of the Standard Model as an effective field theory, with $M$ somewhere in the neighborhood of $10^{16}$ GeV.

Of course, these non-renormalizable terms can be (and in fact, had been) generated in various renormalizable grand-unified theories by integrating out the heavy particles in these theories.  Some calculations in the resulting theories can be assisted by treating them as effective field theories.\footnote{The effective field theories derived by integrating out heavy particles had been considered by T. Appelquist and J. Carrazone, Phys. Rev. D11, 2856 (1975).  In 1980, in a paper titled ``Effective Gauge Theories,'' I used the techniques of effective field theory to evaluate the effects of integrating out the heavy gauge bosons in grand unified theories on the initial conditions for the running of the gauge couplings down to accessible energies: S. Weinberg, Phys. Lett. 91B, 51 (1980).}  But the important  point is that the existence of  suppressed baryon- and lepton-nonconserving terms, and some of their detailed properties, should be expected on much more general grounds, {\em even if the underlying theory is not a quantum field theory at all}.  Indeed, from the 1980s on, it has been increasingly popular to suppose that the theory underlying the Standard Model as well as general relativity is a string theory.

Which brings me to gravitation.  Just as we have learned to live with the fact that there is no renormalizable theory of pion fields that is invariant under the chiral transformation (7), so also we should not despair of applying quantum field theory to gravitation just because  there is no renormalizable theory of the metric tensor that is invariant under general coordinate transformations.  It increasingly seems apparent that the Einstein--Hilbert Lagrangian $ \sqrt{g}R$ is just the least suppressed term in the Lagrangian of an effective field theory containing every possible generally covariant function of the metric and its derivatives.   
  The application of this point of view to long range properties of gravitation has been most thoroughly developed by John Donoghue and his collaborators.\footnote{J. F. Donoghue, Phys. Rev. D50, 3874 (1884); Phys. Lett. 72, 2996 (1994);  lectures presented at the Advanced School on Effective Field Theories (Almunecar, Spain, June 1995), gr-qc/9512024; J. F. Donoghue, B. R. Holstein, B.Garbrecth, and T.Konstandin, Phys. Lett. B529, 132 (2002);  N. E. J. Bjerrum-Bohr, J. F. Donoghue, and B. R. Holstein, Phys. Rev. D68, 084005 (2003).} One consequence of viewing the Einstein--Hilbert Lagrangian as one term in an effective field theory is that any theorem based on conventional general relativity, which declares that under certain initial conditions future singularities are inevitable, must be reinterpreted to mean that under these conditions higher terms in the effective action become important. 

Of course, there is a problem --- the effective theory of gravitation cannot be  used at very high energies, say of the order of the Planck mass, no more than chiral dynamics can be used above a momentum of order $2\pi F_\pi\approx 1$ GeV.  For purposes of the subsequent discussion, it is useful to express this problem in terms of the Wilsonian renormalization group.  The effective action for gravitation takes the form
\begin{eqnarray}
I_{\rm eff} &=& -\int d^4x \;\sqrt{-{\rm Det} g}\Bigg[f_0(\Lambda)+f_1(\Lambda)R \nonumber \\&&+f_{2a}(\Lambda)R^2+f_{2b}(\Lambda)R^{\mu\nu}R_{\mu\nu}\nonumber\\&&+f_{3a}(\Lambda)R^3+\dots\Bigg]\;,
\end{eqnarray}
where here $\Lambda$ is the ultraviolet cutoff, and the $f_n(\Lambda)$ are coupling parameters with a cutoff dependence chosen so that physical quantities are cutoff-independent.
We can replace these coupling parameters  with dimensionless parameters $g_n(\Lambda)$:
\begin{eqnarray}
 &&g_0\equiv \Lambda^{-4}f_0\,;\;g_1\equiv \Lambda^{-2}f_1\,;\;g_{2a}\equiv f_{2a}\,;\;\nonumber\\&&
~~~~g_{2b}\equiv f_{2b}\,;\;g_{3a}\equiv \Lambda^{2}f_{3a}\,;\;\dots ~~~\;.
\end{eqnarray}
Because dimensionless, these parameters must satisfy a renormalization group equation of the form
\begin{equation}
\Lambda \frac{d}{d\Lambda}g_n(\Lambda)=\beta_n\Big(g(\Lambda)\Big)\;.
\end{equation}
In perturbation theory, all but a finite number of the $g_n(\Lambda)$ go to infinity as $\Lambda\rightarrow\infty$, which if true would rule out the use of this theory to calculate anything at very high energy.  There are even examples, like the Landau pole in quantum electrodynamics and the phenomenon of ``triviality'' in scalar field theories, in which the couplings blow up at a {\em finite} value of $\Lambda$.  

It is usually assumed that this explosion of the dimensionless couplings at high energy is irrelevant in the theory of gravitation, just as it is irrelevant in chiral dynamics.  In chiral dynamics, it is understood that at energies of order $2\pi F_\pi\approx m_N$, the appropriate degrees of freedom are no longer pion and nucleon fields, but rather quark and gluon fields.  In the same way, it is usually assumed that in the quantum theory of gravitation, when $\Lambda$ reaches some very high energy, of the order of $10^{15}$ to $10^{18}$ GeV, the appropriate degrees of freedom are no longer the metric and the Standard Model fields, but something very different, perhaps strings.  

But maybe not.  It is just possible that the appropriate degrees of freedom at all energies are the metric and matter fields, including those of the Standard Model.  
The dimensionless couplings can be  protected from blowing up if they are attracted to a finite value $g_{n*}$.  This is known as {\em asymptotic safety.}\footnote{This was first proposed in my 1976 Erice lectures; see footnote 20.}    

Quantum chromodynamics provides an example of asymptotic safety, but one in which the theory at high energies is not only safe from exploding couplings, but also free.  In the more general case of asymptotic safety, the high energy limit $g_{n*}$ is finite, but not commonly zero.

  For asymptotic safety to be possible, it is necessary that all the beta functions should vanish at $g_{n*}$:
\begin{equation}
\beta_n(g_*)=0\;.
\end{equation}
It is also necessary that the physical couplings should be on a trajectory that is attracted to $g_{n*}$.  The number of independent parameters in such a theory equals the dimensionality of the surface, known as the {\em ultraviolet critical surface}, formed by all the trajectories that are attracted to the fixed point.  This dimensionality had better be finite, if the theory is to have any predictive power at high energy.  For an asymptotically safe theory with a finite-dimensional ultraviolet critical surface, the requirement that couplings lie on this surface plays much the same role as the requirment of renormalizability in quantum chromodynamics --- it provides a rational basis for limiting the complexity of the theory.

This dimensionality of the ultraviolet critical surface can be expressed in terms of  the behavior of $\beta_n(g)$ for $g$ near the fixed point $g_*$.  Barring unexpected singularities, in this case we have
\begin{equation}
\beta_n(g)\rightarrow \sum_m B_{nm}(g_m-g_{*m})\;,~~~B_{nm}\equiv \left(\frac{\partial \beta_n(g)}{\partial g_m}\right)_*\;.
\end{equation}
The  solution of Eq.~(16) for $g$ near $g_*$ is then
\begin{equation}
g_n(\Lambda)\rightarrow g_{n*}+\sum_i u_{in}\,\Lambda^{\lambda_i}\;,
\end{equation}
where $\lambda_i$ and $u_{in}$ are the eigenvalues and suitably normalized eigenvectors of $B_{nm}$:
\begin{equation}
\sum_m B_{nm}\,u_{im}=\lambda_i\,u_{in}\;.
\end{equation}
Because $B_{nm}$ is real but not symmetric, the eigenvalues are either real, or come in pairs of complex conjugates.  The dimensionality of the ultraviolet critical surface is therefore equal to the number of eigenvalues of $B_{nm}$ with negative real part.  The condition that the couplings lie on this surface can be regarded as a generalization of the condition that quantum chromodynamics, if it were a fundamental and not merely an effective field theory, would have to involve only renormalizable couplings.

 It may seem unlikely that an infinite matrix like $B_{nm}$ should have only a finite number of eigenvalues with negative real part, but in fact examples of this are quite common.  As we learned from the Wilson--Fisher theory of critical phenomena, when a substance undergoes a second-order phase transition, its parameters are subject to  a renormalization group equation that has a fixed point, with a single infrared-repulsive direction, so that adjustment of a single parameter such as the temperature or the pressure can put the parameters of the theory on an infrared attractive surface of co-dimension one, leading to long-range correlations.  The single infrared-repulsive direction is at the same time a unique ultraviolet-attractive direction, so the ultraviolet critical surface in such a theory is a one-dimensional curve.  Of course, the parameters of the substance on this curve do not really approach a fixed point at very short distances, because at a distance of the order of the interparticle spacing the effective field theory describing the phase transition breaks down.

What about gravitation?  There are indications that here too there is a fixed point, with an ultraviolet critical surface of finite dimensionality.  Fixed points  have been found (of course with  $g_{n*}\neq 0$) using dimensional continuation from $2+\epsilon$ to 4 spacetime dimensions,\footnote{S. Weinberg, in {\em General Relativity}, ed. S. W. Hawking and W. Israel (Cambridge University Press, 1979): 700; H. Kawai, Y. Kitazawa, \& M. Ninomiya, Nucl. Phys. B 404, 684 (1993);   Nucl. Phys. B 467, 313 (1996); T.  Aida \& Y. Kitazawa, Nucl. Phys. B 401, 427 (1997);  M. Niedermaier, Nucl. Phys. B 673, 131 (2003) .}
 by a $1/N$ approximation  (where $N$ is the number of added matter fields),\footnote{L. Smolin, Nucl. Phys. B208, 439 (1982);
 R. Percacci, Phys. Rev. D 73, 041501 (2006).} by lattice methods,\footnote{J. Ambj\o rn, J. Jurkewicz, \& R. Loll, Phys. Rev. Lett. 93, 131301 (2004);  Phys. Rev. Lett. 95, 171301 (2005); Phys. Rev. D72, 064014 (2005);   Phys. Rev. D78, 063544 (2008); and  in {\em Approaches to Quantum Gravity}, ed. D. Or\'{i}ti (Cambridge University Press).} and by use of the truncated exact renormalization group equation,\footnote{M. Reuter, Phys. Rev. D 57, 971 (1998); D. Dou \& R. Percacci, Class. Quant. Grav. 15, 3449 (1998); W. Souma, Prog. Theor. Phys. 102, 181 (1999); O. Lauscher \& M. Reuter, Phys. Rev. D 65, 025013 (2001); Class. Quant. Grav. 19. 483 (2002);  M. Reuter \& F. Saueressig, Phys Rev. D 65, 065016  (2002); O. Lauscher \& M. Reuter, Int. J. Mod. Phys. A 17, 993 (2002);  Phys. Rev. D 66, 025026 (2002); M. Reuter and F. Saueressig, Phys Rev. D 66, 125001 (2002); R. Percacci \& D. Perini, Phys. Rev. D 67, 081503 (2002);  Phys. Rev. D 68, 044018 (2003); D. Perini, Nucl. Phys. Proc. Suppl. C 127, 185 (2004); D. F. Litim, Phys. Rev. Lett. {\bf 92}, 201301 (2004); A. Codello \& R. Percacci, Phys. Rev. Lett. 97, 221301 (2006); A. Codello, R. Percacci, \& C. Rahmede, Int. J. Mod. Phys. A23, 143 (2008);  M. Reuter \& F. Saueressig, 0708.1317; P. F. Machado and F. Saueressig, Phys. Rev. D77, 124045 (2008); A. Codello, R. Percacci, \& C. Rahmede, Ann. Phys. 324, 414 (2009);  A. Codello \& R. Percacci, 0810.0715; D. F. Litim 0810.3675; H. Gies \& M. M. Scherer, 0901.2459; D. Benedetti, P. F. Machado, \& F. Saueressig, 0901.2984, 0902.4630; M. Reuter \& H. Weyer, 0903.2971.} initiated in 1998 by Martin Reuter.  In the last method,  which had earlier been  introduced in condensed matter physics\footnote{F. J. Wegner and A. Houghton, Phys. Rev. A8, 401 (1973).} and then carried over to particle  theory,\footnote{J. Polchinski, Nucl. Phys. B231, 269 (1984); C. Wetterich, Phys. Lett. B 301, 90 (1993).} one derives an exact renormalization group equation for the total vacuum amplitude $\Gamma[g,\Lambda]$ in the presence of a background metric $g_{\mu\nu}$  with an {\em infrared} cutoff $\Lambda$.  This is the action to be used in calculations of the true vacuum amplitude in calculations of graphs with an {\em ultraviolet} cutoff $\Lambda$. To have equations that can be solved, it is necessary to truncate these renormalization group equations, writing  $\Gamma[g,\Lambda]$ as a sum of just a finite number of terms like those shown explicitly in Eq.~(14), and ignoring the fact that the beta function inevitably does not vanish for the couplings of other terms in $\Gamma[g,\Lambda]$ that  in the given truncation are assumed to vanish.                                                                                                                                                                                                                                                              

Initially only two terms were included in the truncation of $\Gamma[g,\Lambda]$ (a cosmological constant and the Einstein--Hilbert term $\sqrt{g}R$), and a fixed point was found with two eigenvalues $\lambda_i$, a pair of complex conjugates with negative real part.  Then a third operator ($R_{\mu\nu}R^{\mu\nu}$ or the equivalent) was added, and a third eigenvalue was found, with $\lambda_i$ real and negative.  This was not encouraging.  If each time that   new terms were included in the truncation, new eigenvalues appeared with negative real  part, then the ultraviolet critical surface would be infinite dimensional, and the theory, though free of couplings that exploded at high energy, would lose all predictive value at high energy.

In just the last few years calculations have been done that allow  more optimism.  Codello, Percacci, and Rahmede\footnote{A. Codello, R. Percacci, \& C. Rahmede, Int. J. Mod. Phys. A23, 143 (2008)} have considered a Lagrangian containing all terms $\sqrt{g}R^n$ with $n$ running from zero to a maximum value $n_{\rm max}$, and find that the ultraviolet critical surface has dimensionality 3 even when $n_{\rm max}$ exceeds 2, up to the highest value $n_{\rm max}=6$ that they considered, for which the space of coupling constants is 7-dimensional.  Furthermore,  the three eigenvalues they find with negative real part seem to converge as $n_{\rm max}$ increases, as shown in the following table of ultraviolet-attractive eigenvalues:
$$
\begin{array}{ccccc}
n_{\rm max}=2: &~~~~~& -1.38\pm 2.32i &~~~~~& -26.8 \\
n_{\rm max}=3: &~~~~~& -2.71\pm 2.27i &~~~~~& -2.07 \\
n_{\rm max}=4: &~~~~~& -2.86 \pm 2.45i &~~~~~& -1.55 \\
n_{\rm max}=5: &~~~~~& -2.53 \pm 2.69i &~~~~~& -1.78  \\
n_{\rm max}=6: &~~~~~& -2.41 \pm 2.42i &~~~~~& -1.50
\end{array}
$$
In a subsequent paper\footnote{A. Codello, R. Percacci, \& C. Rahmede, Ann. Phys. 324, 414 (2009)} they added matter fields, and again found just three ultraviolet-attractive eigenvalues.  Further, this  year Benedetti, Machado, and Saueressig\footnote{D. Benedetti, P. F. Machado, \& F. Saueressig, 0901.2984, 0902.4630} considered a truncation with a different four terms,  terms proportional to $\sqrt{g}R^n$ with $n=0, 1$ and 2 and also $\sqrt{g}C_{\mu\nu\rho\sigma}C^{\mu\nu\rho\sigma}$ (where $C_{\mu\nu\rho\sigma}$ is the Weyl tensor) and they too find just three ultraviolet-attractive eigenvalues, also when matter is added.  If this pattern of eigenvalues continues to hold in future calculations, it will begin to look as if there is a quantum field theory of gravitation that is well-defined at all energies, and that has just three free parameters.

The natural arena for application of these ideas is in the physics of gravitation at small distance scales and high energy --- specifically, in the early universe.  A start in this direction has been made by Reuter and his collaborators,\footnote{ A. Bonanno and M. Reuter, Phys. Rev. D 65, 043508 (2002); Phys. Lett. B527, 9 (2002); M.  Reuter and F. Saueressig, J. Cosm. and Astropart. Phys. 09, 012 (2005).}
 but much remains to be done.

I am grateful for correspondence  about recent work on asymptotic safety with D. Benedetti, D. Litim, R. Percacci, and M. Reuter, and to G. Colangelo and J. Gasser for inviting me to give this talk.  This material is based in part on work supported by the National Science Foundation under Grant NO. PHY-0455649 and with support from The Robert A. Welch Foundation, Grant No. F-0014.

\end{document}